# Scientific Impact of Graph-Based Approaches in Deep Learning Studies - A Bibliometric Comparison

**Ilker TURKER [1*], Serhat Orkun TAN [2]**

[1] Karabuk University, Dept. of Computer Engineering, Karabuk, Turkey (iturker@karabuk.edu.tr)
[2] Karabuk University, Dept. of Electrical Engineering, Karabuk, Turkey

## ABSTRACT

Applying graph-based approaches in deep learning receives more attention over time. This study presents statistical analysis on the use of graph-based approaches in deep learning and examines the scientific impact of the related articles. Processing the data obtained from the Web of Science database, metrics such as the type of the articles, funding availability, indexing type, annual average number of citations and the number of access were analyzed to quantitatively reveal the effects on the scientific audience.

It's outlined that deep learning-based studies gained momentum after year 2013, and the rate of graph-based approaches in all deep learning studies increased linearly from 1% to 4% within the following 10 years. Conference publications scanned in the Conference Proceeding Citation Index (CPCI) on the graph-based approaches receive significantly more citations. The citation counts of the SCI-Expanded and Emerging SCI indexed publications of the two streams are close to each other. While the citation performances of the supported and unsupported publications of the two sides were similar, pure deep learning studies received more citations on the journal publication side and graph-based approaches received more citations on the conference side. Despite their similar performance in recent years, graph-based studies show twice more citation performance as they get older, compared to traditional approaches. Annual average citation performance per article for all deep learning studies is 11.051 in 2014, while it is 22.483 for graph-based studies. Also, despite receiving 16% more access, graph-based papers get almost the same overall citation over time with the pure counterpart. This is an indication that graph-based approaches need a greater bunch of attention to follow, while pure deep learning counterpart is relatively simpler to get inside.

## 1. INTRODUCTION

Deep Neural Networks (DNNs), a form of Artificial Neural Networks (ANNs) with complexity in network structure consisting of cascaded chain across many hidden layers, has emerged as a principal component of Artificial Intelligence (AI) ecosystem. It takes benefit of an imitative learning principal from biological systems by changing the distribution of weights with the capability of learning patterns in the input [1]. Later abbreviated as Deep Learning (DL), these models successfully handle learning tasks related to classification, clustering and regression problems [2], including many fields such as medicine [3, 4], biomedical science [5, 6], telecommunications [7], cyber-security [8], finance [9], mechanics [10], machining [11], material design [12] etc.

The success of DNN approach beyond expectation, many topologic variations of DL architectures have emerged during the last decade [13]. Deep Belief Networks (DBN), Restricted Boltzman Machines (RBM), Convolutional Neural Networks (CNN), Recurrent Neural Networks (RNN), Gated Recurrent Units (GRU), Long-Short Term Memory Networks (LSTM), Autoencoders (AE), Generative Adversarial Networks (GAN) are a bunch of these variations, each of which has capabilities of handling specific types of data [14]. DL techniques can significantly output higher performance compared to traditional Machine Learning (ML) methods [15], thanks to its more capacitive structure.

Recent years also witnessed a noteworthy interest in data representation techniques, presenting various approaches to reconstruct data with higher order representations compared to traditional feature learning approaches. Many data representation learning methods have been proposed including groundbreaking samples as principal component analysis (PCA), linear discriminant analysis (LDA) and generalized discriminant analysis (GDA). Later in DL era, some embedding techniques have been preferred to reduce the dimensionality, especially for graph structured data [16]. These approaches form the later end of data transformation spectrum, after the graph-based representations have been popular for a variety of data types.

Graph-structured data, built as a relational representation from raw data, attracts increasing attention recently. Research on graph representation learning also induced generation of techniques for deep graph embeddings and adaptation of convolutional neural networks (CNNs) to graph-structured data. It also facilitated development of specific deep learners with the capability to learn from this kind of data. These expansion in graph representation direction resulted in better success levels in numerous domains including 3D vision, network analysis and classification, recommender systems, time series classification etc. [17, 18].

This study aims to present a statistical view on the use of graph-based approaches in deep learning in the literature, also examining the scientific impact of the articles. In the light of the bibliometric data obtained from the Web of Science (WoS), metrics such as the type of the articles, funding availability, indexing type, number of citations and reads were analyzed in order to quantitatively output the effects of the related articles on the scientific world. A comparative analysis between graph-based and classical approaches in deep learning tasks is driven to assess the scientific value or attractiveness of graph representation methods.

## 2. DATA AND METHODS

WoS Core Collection enables access to bibliometric data through the user interface enabling querying and listing publications. Moreover, the system enables batch downloading the results in form of 1000-record bins. For the queries resulting in hundreds of thousands, a hand-driven task to download the whole data gets very difficult. Therefore, automated downloading of a big data collection can be facilitated by browser macros. iMacros for Chrome has been our choice to access the data collection for this study, which is conducted over a collection of ~150,000 records. The provided results include many attributes of the publications including publication type, years, research field, authors, affiliations, journal or conference titles, publishers, WoS index etc. We performed queries for the two counterparts of the comparative study as:

Deep Learning (General): TS=("deep learning") and PY=(2000-2023)

Deep Learning (Graph structured): TS=("deep learning" AND "graph") and PY=(2000-2023)

Retrieved results accordingly include records within a time span of 2000-2023.

A statistical outline on a variety of available fields of data is performed, including publication years, document types, indexing information, funding availability etc. After providing a brief view of data for the mentioned basic parameters, analysis related with the performance metrics such as citation count and usage count are presented with respect to the above-mentioned parameters. Analysis also involves in statistical inferences from probability distributions of the mentioned metrics.

## 3. RESULTS AND DISCUSSION

### 3.1 Basic Statistical Inferences

This section evaluates comparisons of basic bibliometric statistics through available attributes from the WoS database. Together with the population or percentage-based graphics for both unrestricted and graph-based approaches of deep learning studies, we also present the evolution of tendency of graph-based studies in comparison with all studies by calculating a $t_g$ (*tendency of graph-based methods)* percentage metric calculated as below:

$$t_g = \frac{100(p_g - p_t)}{p_t} \quad (1)$$

Where $p_g$ and $p_t$ are percentages for graph-based and traditional approaches for the given interval or value.

#### 3.1.1 Publication Count

We present evolution of number of publications for years for both types of studies in as Fig.1. To briefly explain, $t_g$ value for year 2019 is calculated as following: $p_g(2019)$ is the ratio of graph-based publications published in 2019 in all graph-based publications (found as 11.99%). Similarly, $p_t(2019)$ is calculated as the ratio of deep learning publications in 2019 to all collection (found as 15.24%). Then using Eq.1, $t_g$ is calculated as -21.29%, with a meaning that graph-based studies' count has a disadvantage of 21.29% compared to all deep learning studies for that year.

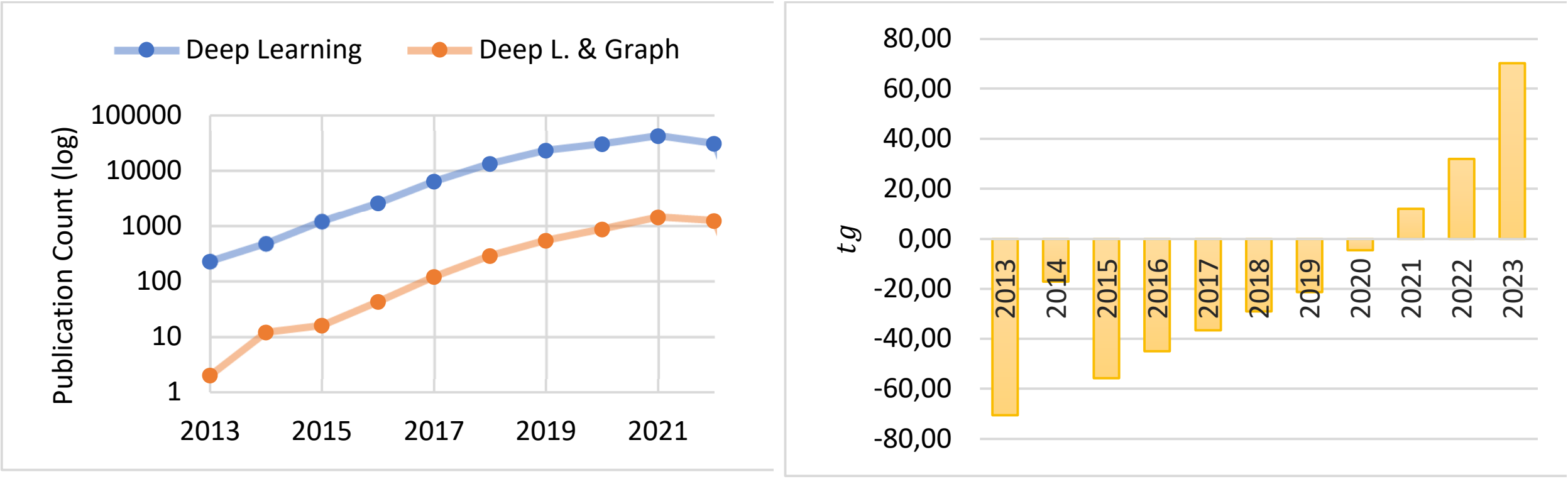

(a) (b)

**Fig.1. a) Publication count vs. years in logarithmic scale. b) Tendency of graph-based approaches ($t_g$) vs. years.**

*Unrestricted DL studies dataset consists of 146,328 records, while graph-induced counterpart includes 4,338 records (~3%), which is also included by the former.*

Results in Fig.1 indicate that both streams have exponentially increasing trends, while increasing rate for graph-based approaches is greater for recent years. We also present Fig.2 to show the ratio of graph-based approaches to all deep-learning studies in yearly representation. This figure indicates a linear increase in preferring graph-based approaches by the researchers.

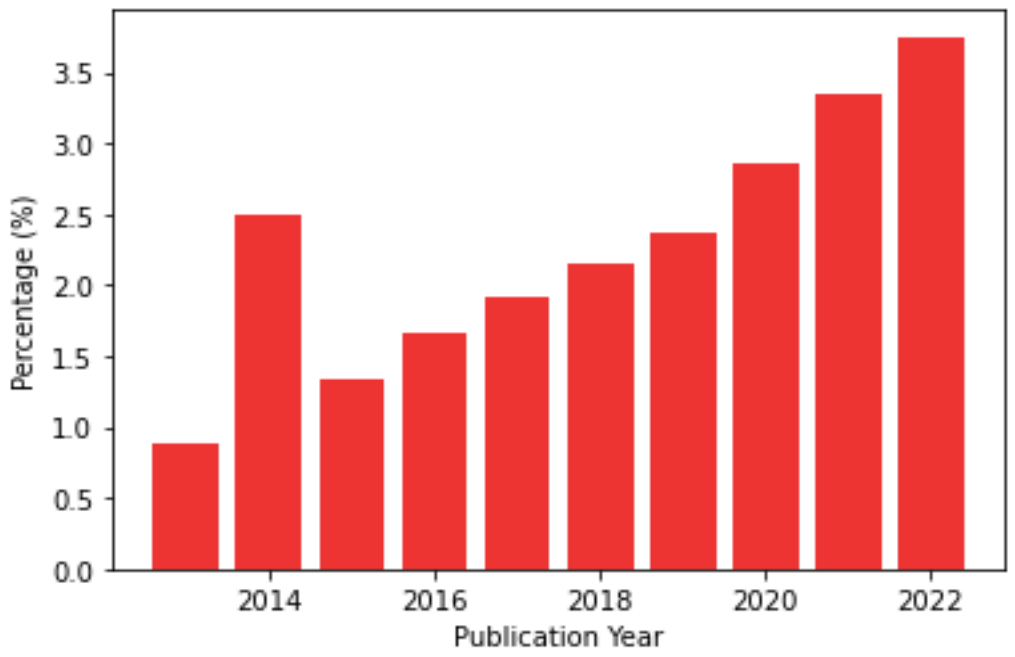

**Fig. 2. Evolution of percentage of graph-based studies in all deep learning studies.**

### 3.1.2 Document Types

Publications in WoS database are subdivided into document types as articles, conference papers, reviews and so on. Dispersion of data according to this attribute is given in Fig.3, where percentages are calculated locally for each stream. This figure indicates that graph-based studies are more published as articles, proceeding papers and early-access (journal) papers, while unrestricted deep learning studies have greater percentage for review papers.

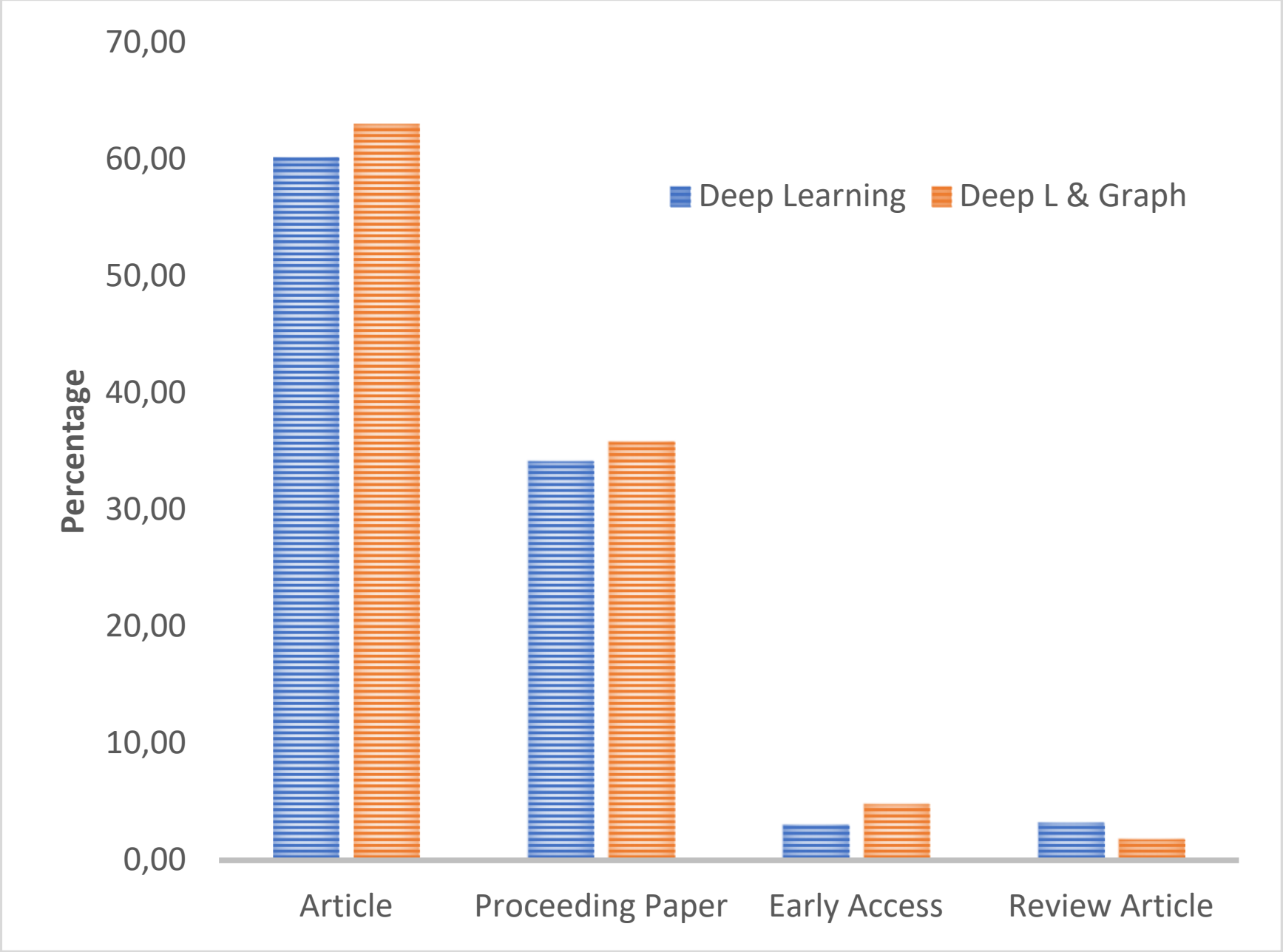

**Fig.3. Distribution of publications according to document types.**

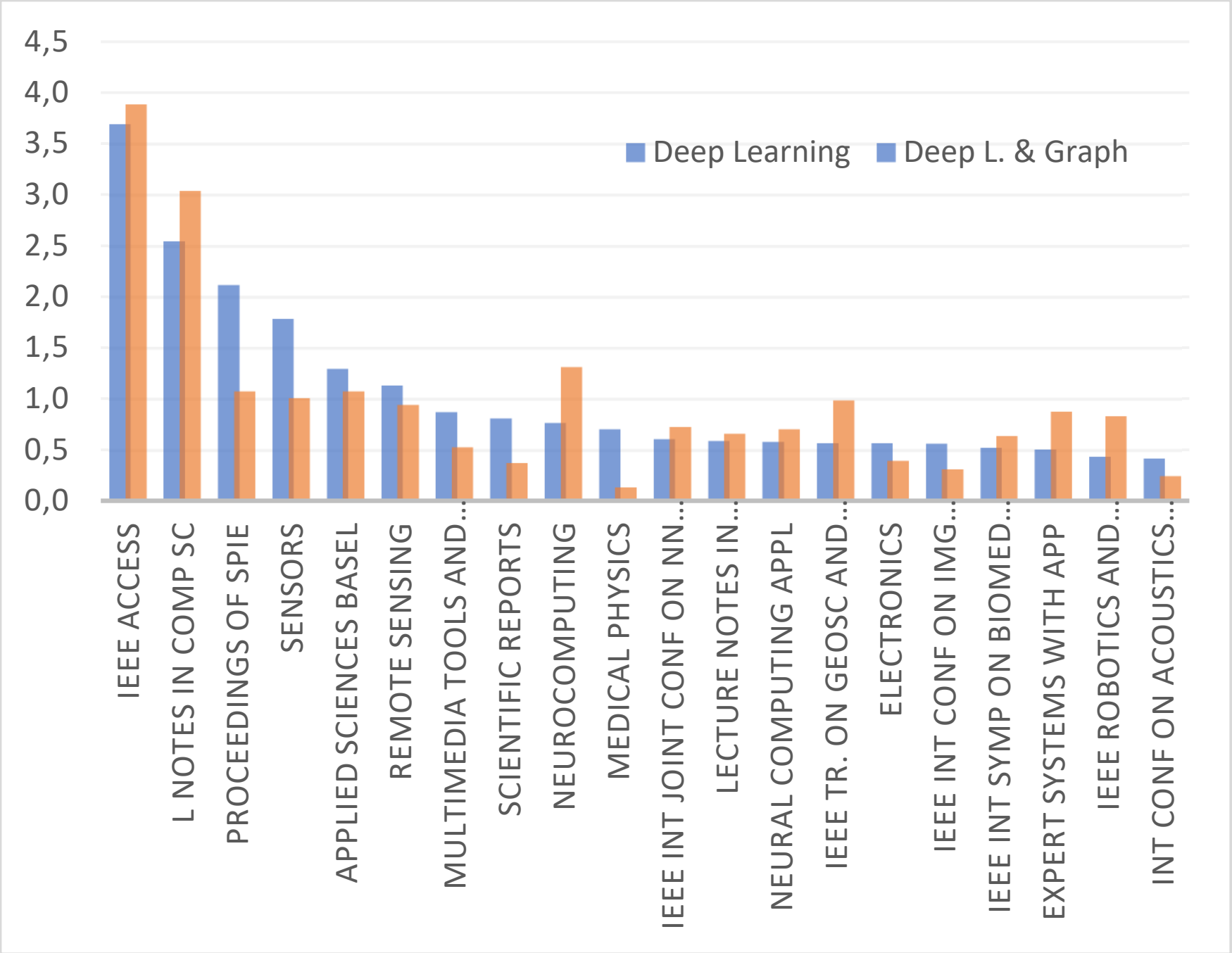

**Fig. 4. Percentage of publications for publication titles (sources).**

### 3.1.3 Publication Titles (Sources)

Publication titles or sources is an important consideration in academy, having underpinnings in research preferences of the societies they belong to. Publication ratios with respect to the sources including the top 20 titles according to publication count is given in Fig. 4. The figure indicates that some traditional and high-reputation conferences and related proceeding series (such as SPIE, Lecture Notes in Computer Science) have a critical role in publishing in deep learning field. IEEE Access takes the massive load of publication above 3.5% of whole, while the closest volume is satisfied by Sensors journal with half percentage compared to IEEE Access. First two sources (IEEE Access and Lecture Notes in Computer Science) seem to selectively publish graph-based deep learning studies, while this positive discrimination is also visible for some journals including Neurocomputing, IEEE Transactions on Geoscience and Remote Sensing, Expert Systems with Applications, IEEE Robotics and Automation Letters. Tendency ($t_g$) plots for these sources are also provided in Fig. 5, indicating positively (left end) and negatively (right end) tendencies for publishing graph-based DL papers.

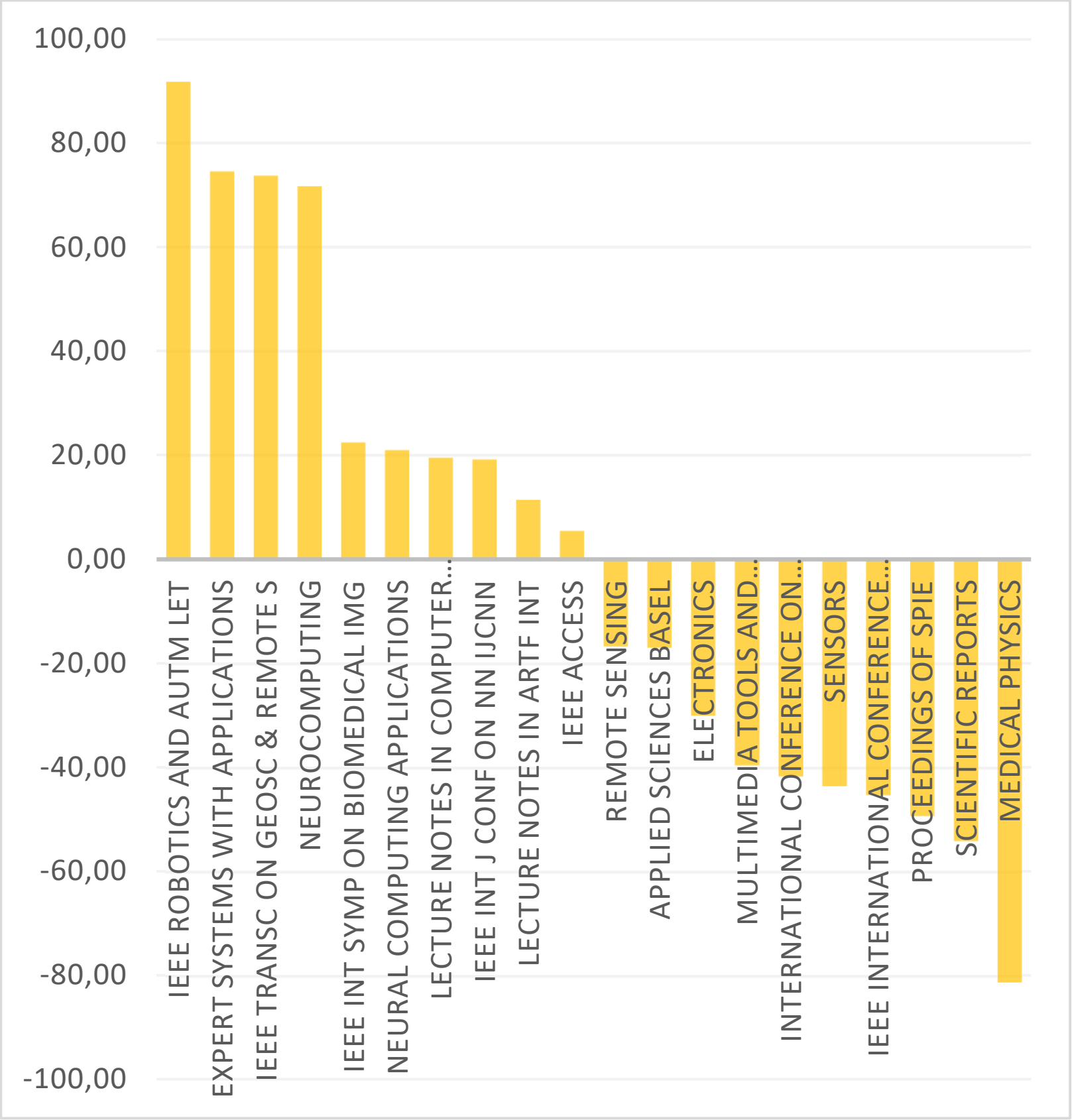

**Fig. 5. Tendency of publishing graph-based studies ($t_g$) wrt. publication sources.**

### 3.1.4. Research Areas

DL has attracted the attention of various scientific fields. Mostly being related with computer science, any other field of science has considerable deal with it. Fig. 6 presents distribution of this deal for several research areas, along with the graph-induced tendency metric $t_g$.

Among the research areas dealing with DL techniques, biochemistry, computational biology, chemistry, mathematics, computer science and remote sensing fields significantly promote usage of graph-based approaches. These are the most related fields with network science, being close to express entities in graph representations. Therefore, this outcome fits our expectations originating from network science. Remaining evaluations from this figure are left to the readers' attention.

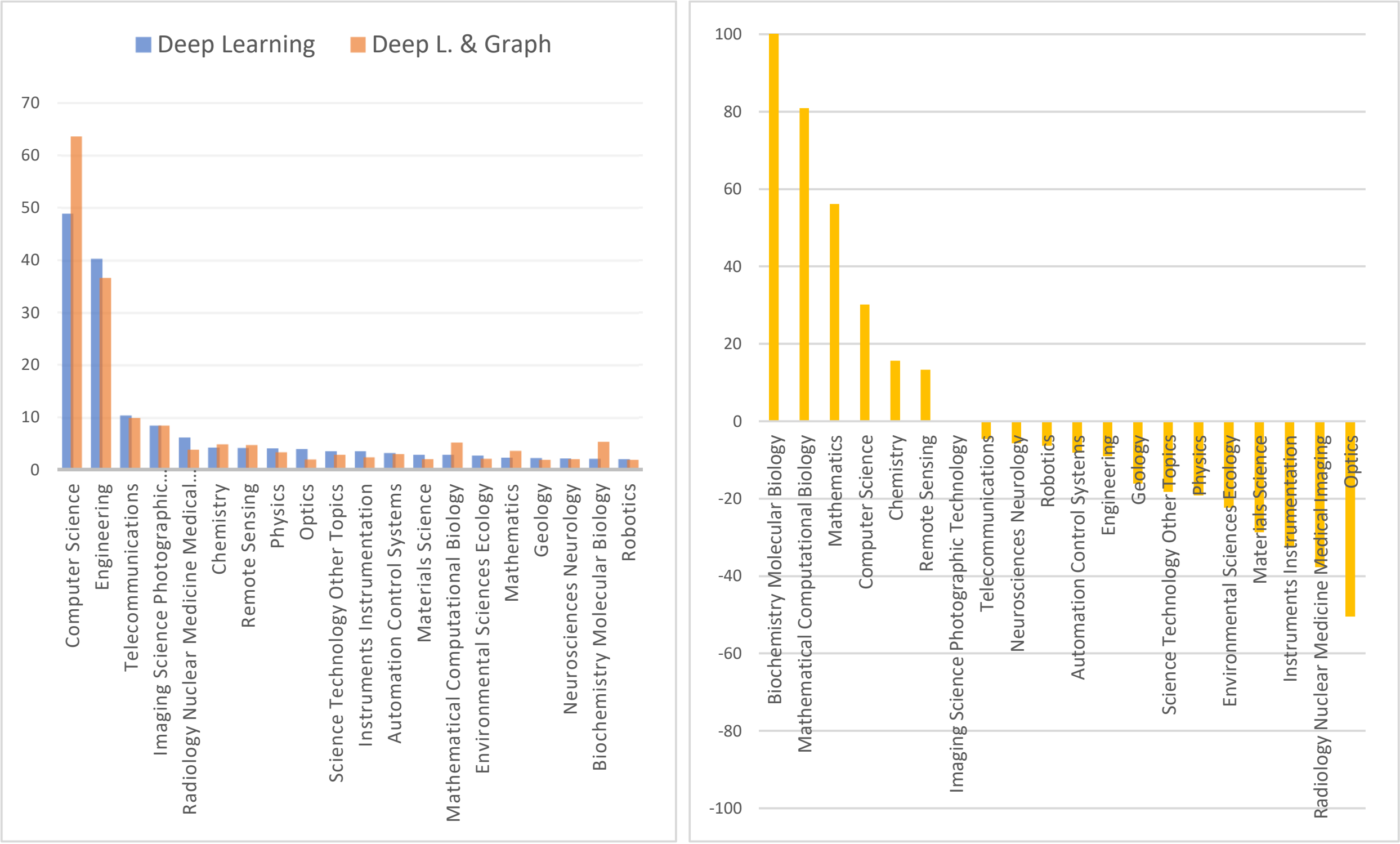

(a) (b)

**Fig. 6. a) Dispersion for Research Areas. b) Tendency of publishing graph-based studies with respect to research areas ($t_g$).**

### 3.1.5 Countries

Dispersion of scientific interest to DL and graph-ed DL techniques with respect to countries and regions is an interesting outcome of this study. We first present a map visualization of scientific interest to DL techniques in Fig. 7, dismissing its graph-related counterpart due to the difficulty to recognize small differences in colors. This figure clearly shows that Republic of China (35% of all studies) is dominating the DL related studies, while USA (21.9%), India (7.3%), South Korea (6%), England (5.3%) and Germany (4.3%) are the following 5 countries with highest rates.

We also present Fig. 8 to analyze the differences of unrestricted vs. graph-based studies in DL domain. This figure indicates that countries having significantly higher scientific interest in graph-based techniques in DL are Rep. of China, Singapore, Australia, Canada and USA. Contrarily, countries including India, Taiwan,

Netherlands, Brazil, Turkey are significantly in negative side to use graph-based approaches. Further evaluations would be possible if additional information was available for these countries.

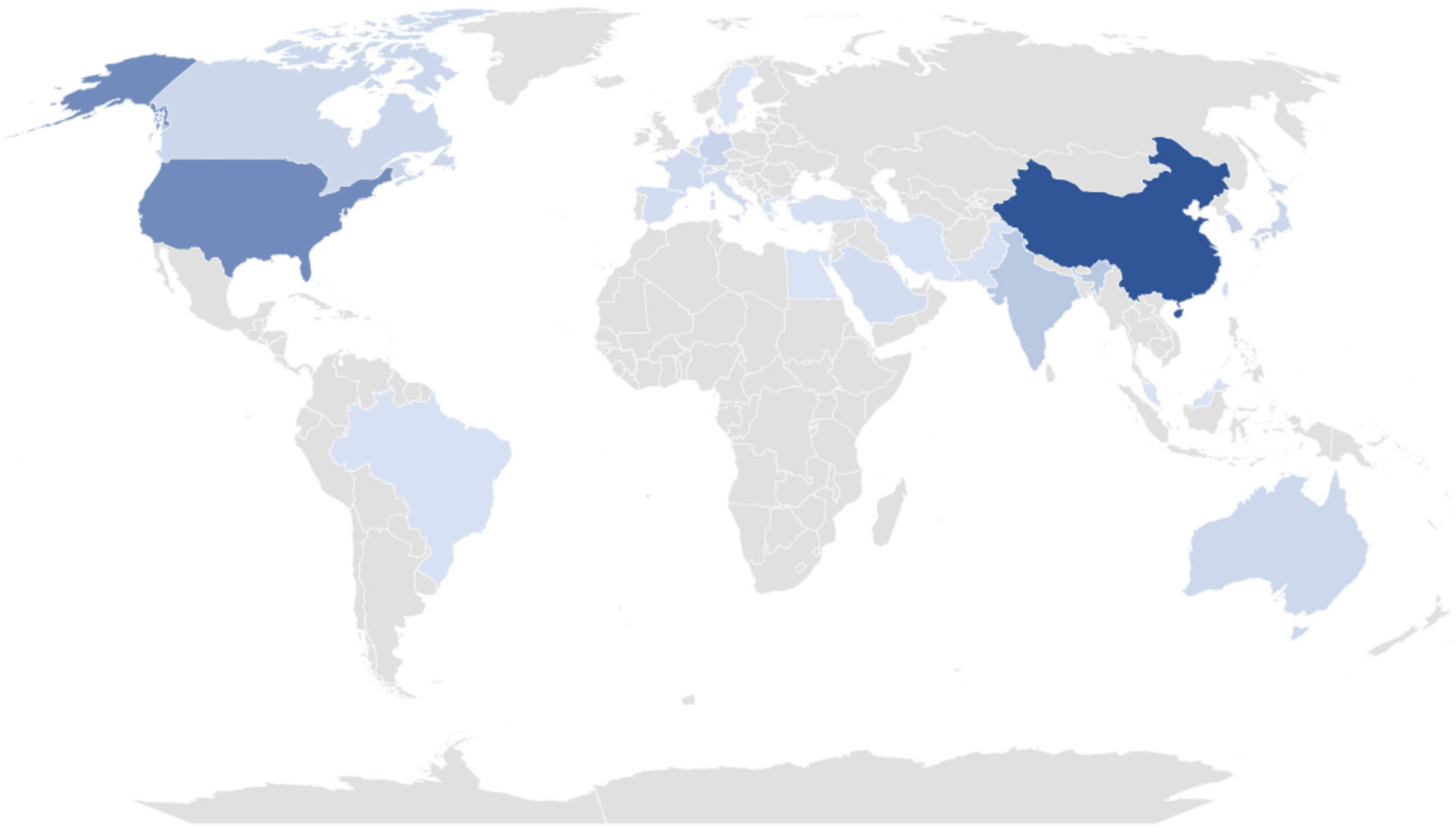

**Fig. 7. Visualization of DL-related paper count from countries/regions of the world.**

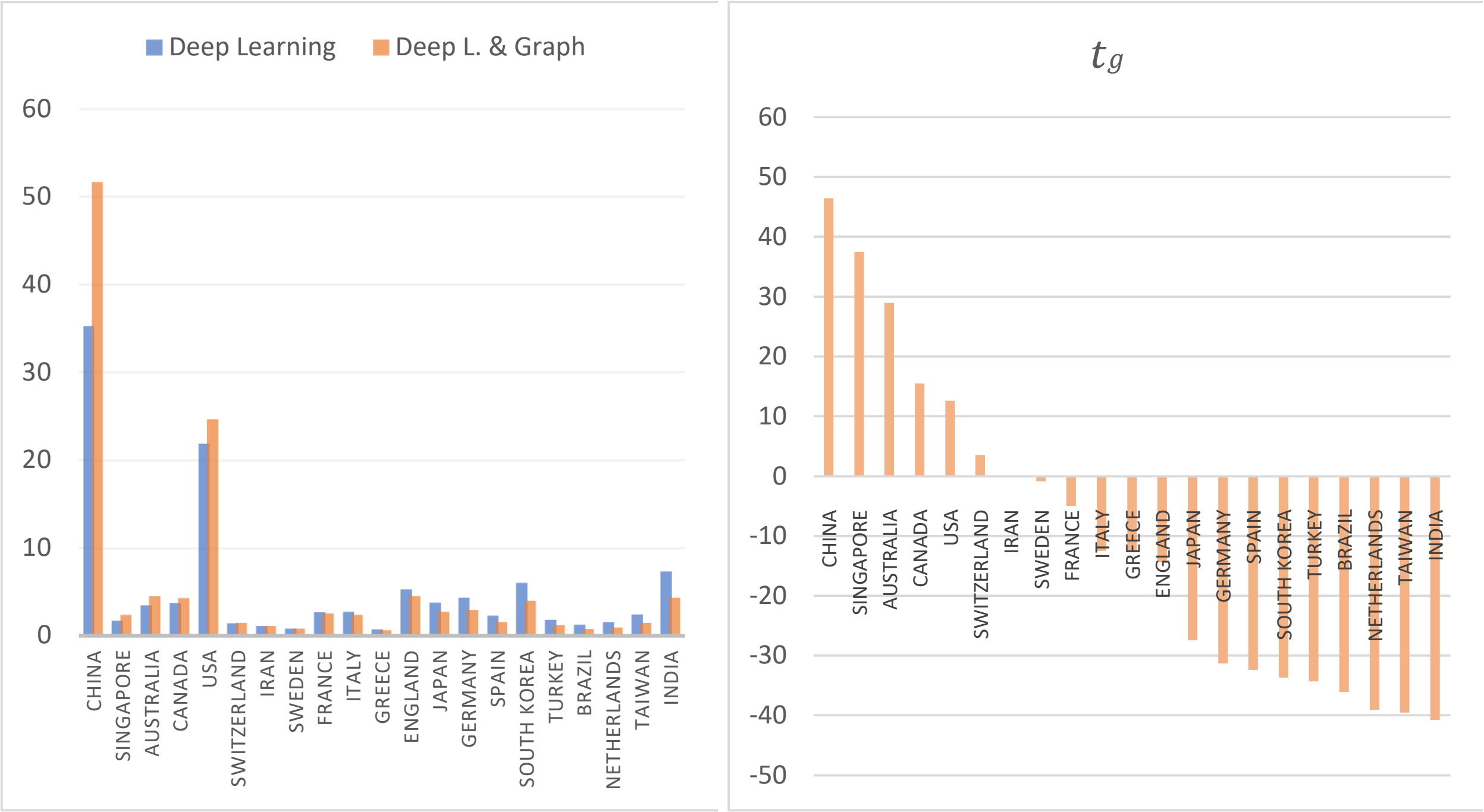

(a) (b)

**Fig. 8. a) Dispersion for countries. b) Tendency of publishing graph-based studies for countries ($t_g$).**

### 3.1.6 WoS Index

The last evaluation available in this subsection is the indexing type. Authors strictly consider this attribute because it is mainly related with scientific reputation or getting some academic degrees. The data related to indexing is visualized in Fig. 9 together with its $t_g$ metric. Left panel of Fig. 9 indicate that studies from both streams are approximately 60% of SCI-Expanded indexed, followed by Conference Proceedings Citation Index – Science (CPCI-S) of ~35%. Remaining studies are indexed in Emerging Science Citation Index (ESCI), Social Sciences Citation Index (SSCI) and others with lowest rates. A noteworthy outcome of this figure is, researchers prefer publishing conference papers at a noticeably high rate in this scope, which requires less time and effort compared to ESCI or BKCI indexed publishing, to facilitate fast and easy access to their studies. This tendency mostly originates from the perception that technology in this field is in a rapid development, which means you have to publish as soon as possible to be recognized and to take advantage of a possible impact of a current research.

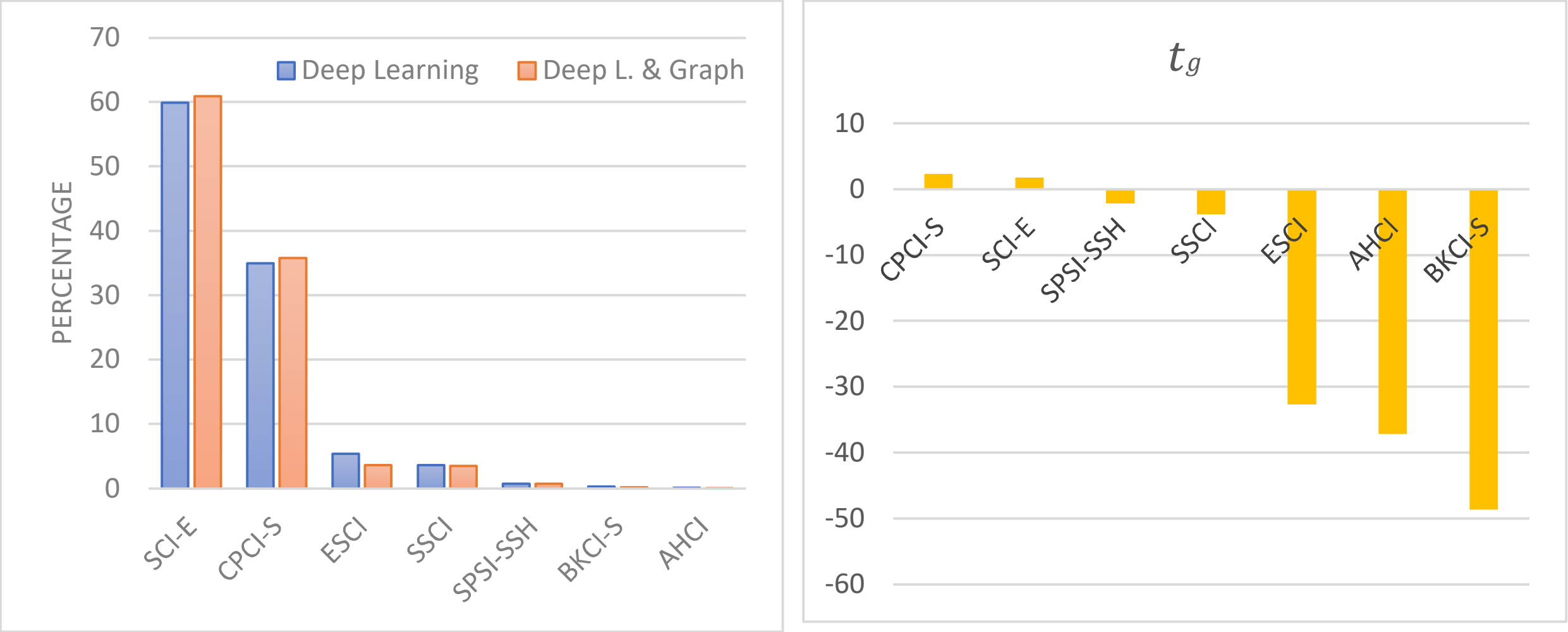

**Fig. 9. a) Dispersion for indexing type. b) Tendency of publishing graph-based studies wrt. indexing type ($t_g$).**

The significant outcome from $t_g$ plot is, SSCI, ESCI, AHCI (Arts and Humanities Citation Index) and BKCI-S (Book Citation Index - Science) indexed studies are strongly in negative side to use graph-based approaches, a list including the indexes farthest to the AI domain, matching our expectation. An overall conclusion from this section may be, graph-induced learning is in the most recent and enhanced side of the AI field.

### 3.2 Evaluating Scientific Impact

While the previous section evaluates the tendency of publishing graph-based studies in DL scope in conjunction with publication types and titles, research areas, countries, WoS index, and yearly evolution of publication counts, this section focuses on the differences in scientific impact, in case of graph-based approaches are followed. The availability of fields in WoS dataset related with citation and usage count along with the previously mentioned attributes enables extraction of impact-based inferences.

First, citation count is a yearly increasing attribute therefore, we normalized it dividing by the age of the publication. As a result, we generated a new attribute as *"normalized citation count"*, later is used in presentation of the results. We label this data as **"yearly normalized citations"** as the horizontal axis of probability distributions of related subsections below. This attribute is also aggregated with respect to years, leading a yearly plot of evolution of citation metrics.

#### 3.2.1. Average Normalized Citation Count vs. Years

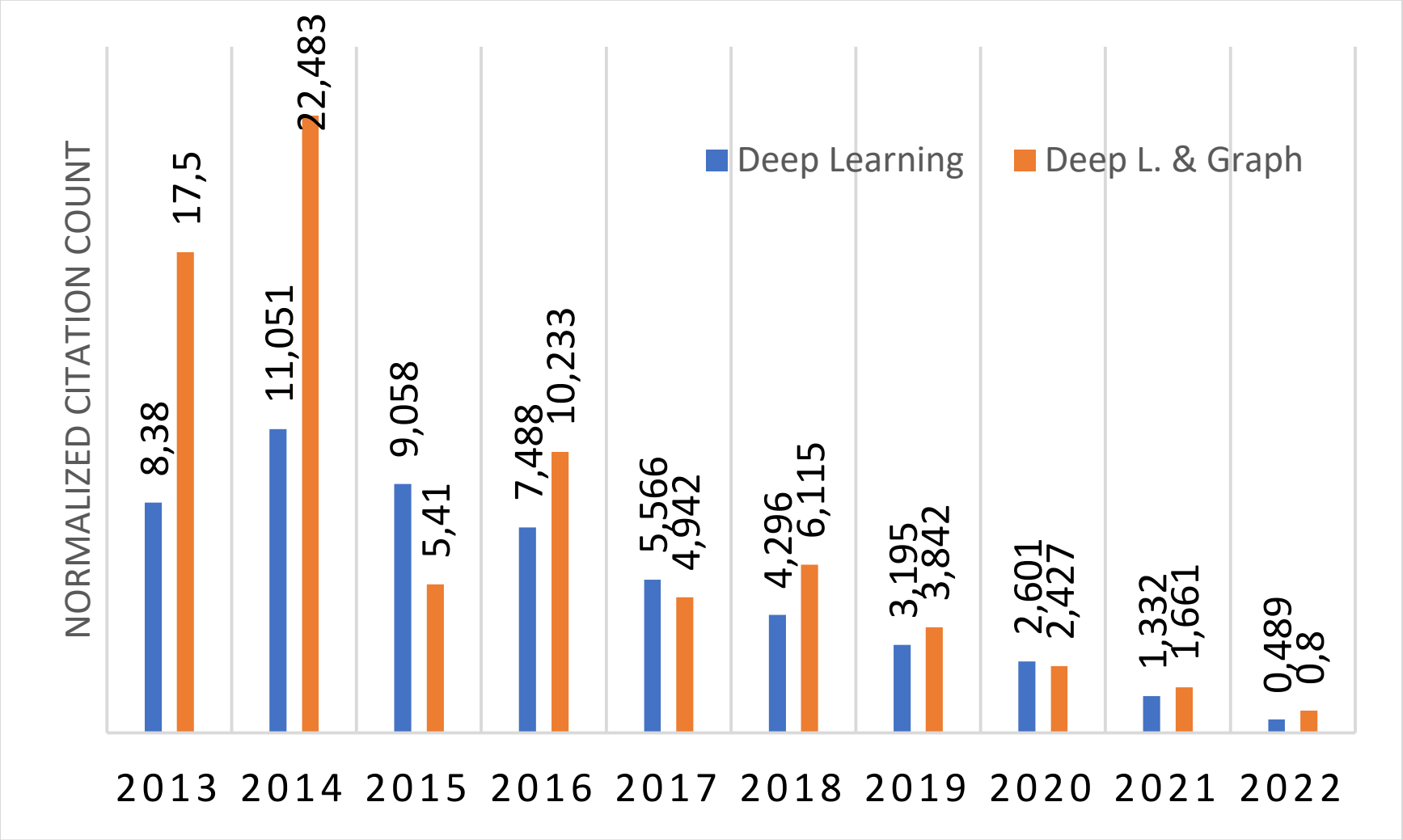

**Fig. 10. Average normalized citation count presented yearly, for unrestricted deep learning and graph-induced studies separately.**

Dispersion of average normalized citations for years is presented in Fig. 10. Despite being normalized over years, average citation count tends to increase with the age of the publications. This indicates that citation performance of papers increases with a super-linear trend for years, eventually growing faster than a linear relation. Effect of aging on citation performance is significantly greater for graph-based studies, especially for papers published in 2013 and 2014.

#### 3.2.2. Distribution of Average Normalized Citation Count for Years

For all distribution plots given in this section, left panel stands for unrestricted DL studies, while right panel stands for graph-based studies.

**i. General Distribution:** This subsection presents general distribution for normalized citation counts (from WoS collection and all databases) together with normalized usage (download) count distributions given in the same plot. Distribution plots given in Fig. 11 show that although unrestricted DL and graph-based studies have comparable citation and usage distributions in the left panel, right panel demonstrates noticeably lower probabilities for citation compared to usage. This indicates that convertibility of reading to citation is lower in graph side, compared to unrestricted side.

As given in Fig. 12, average normalized citation counts are about 20% higher for SCIE indexed DL studies, whereas this number is significantly (67%) greater for CPCI indexed graph-based studies. In BKCI indexed side, unrestricted DL studies have almost 6 times greater citation compared to graph side. To conclude, unrestricted DL studies indexed in SCIE or BKCI are favorably cited more, while CPCI indexed graph-based studies are advantageous in citation potential.

(a) Deep Learning | (b) Deep L. & Graph

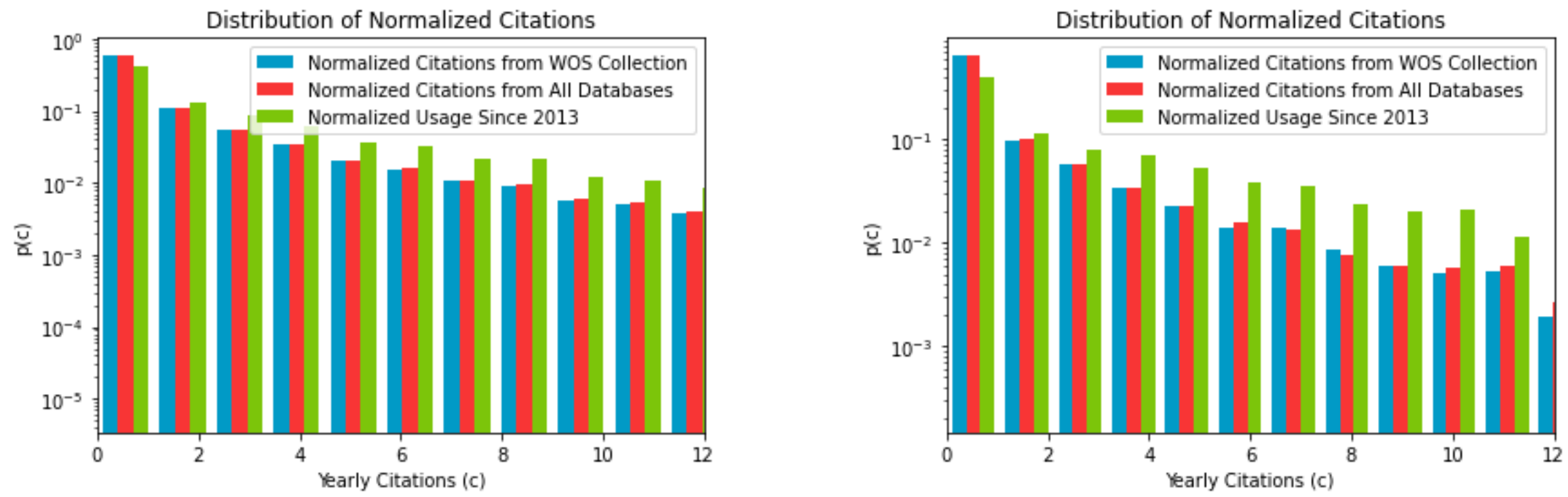

| (a) Deep Learning | | (b) Deep L. & Graph | |
|---|---|---|---|
| Avg. Norm. Citation - WOS Collection | : **2.357** | Avg. Norm. Citation - WOS Collection | : **2.359** |
| Avg. Norm. Citation - All Databases | : **2.423** | Avg. Norm. Citation - All Databases | : **2.429** |
| Avg. Norm. Usage Since 2013 | : **3.596** | Avg. Norm. Usage Since 2013 | : **4.169** |

**Fig. 11. Distribution for normalized citation counts from WoS collection and all databases, together with normalized usage since 2013. Average values are also provided below the panels.**

This fact is also noticeable in averaged values given below the panel. Despite having significantly greater average usage count, graph-based approaches have almost equal average normalized citation to the unrestricted studies.

### ii. Indexing Type

(a) Deep Learning | (b) Deep L. & Graph

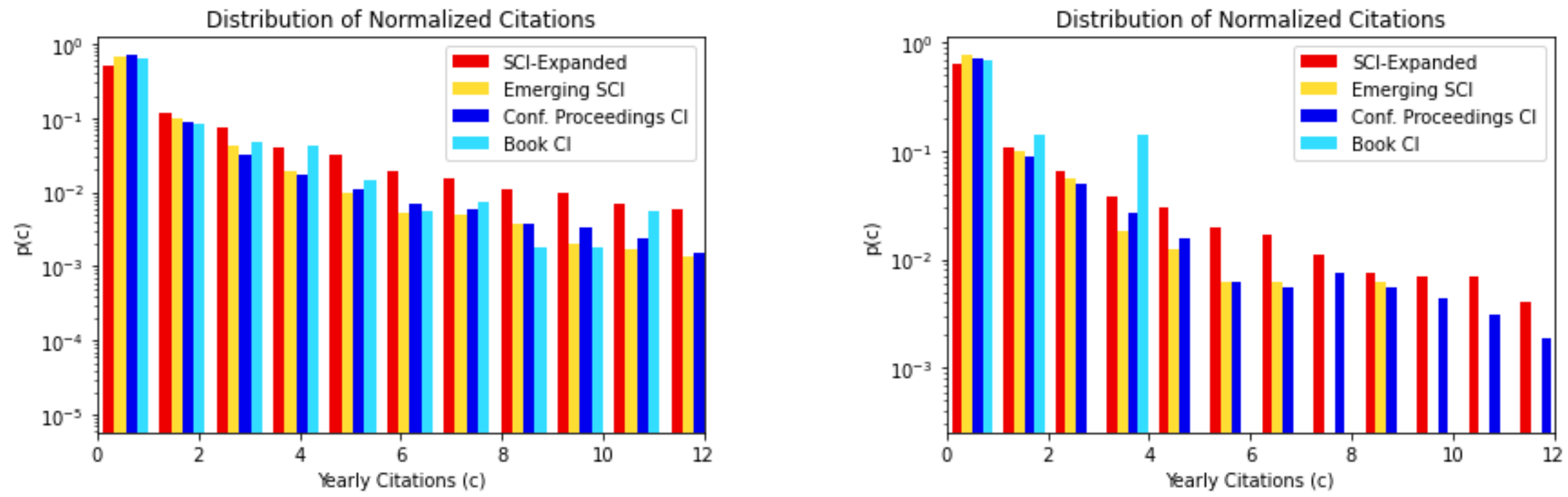

| (a) Deep Learning | | (b) Deep L. & Graph | |
|---|---|---|---|
| Avg. Normalized Citation - SCIE | : **2.987** | Avg. Normalized Citation - SCIE | : **2.459** |
| Avg. Normalized Citation - ESCI | : **1.019** | Avg. Normalized Citation - ESCI | : **1.049** |
| Avg. Normalized Citation - CPCI | : **1.417** | Avg. Normalized Citation - CPCI | : **2.356** |
| Avg. Normalized Citation - BKCI | : **5.919** | Avg. Normalized Citation - BKCI | : **1.057** |

**Fig. 12. Distribution for normalized citation counts, separately given for four indexing services. Average values are also provided below the panels.**

The interest in graph side distinctly condenses at conference publications. Impact of CPCI-indexed studies in graph-based side is surprisingly close to SCIE-indexed studies. SCIE and CPCI-indexed publications attract 1.5 to 3 times more attention compared to ESCI-indexed publications.

**iii. Funding Availability**

(a) Deep Learning | (b) Deep L. & Graph

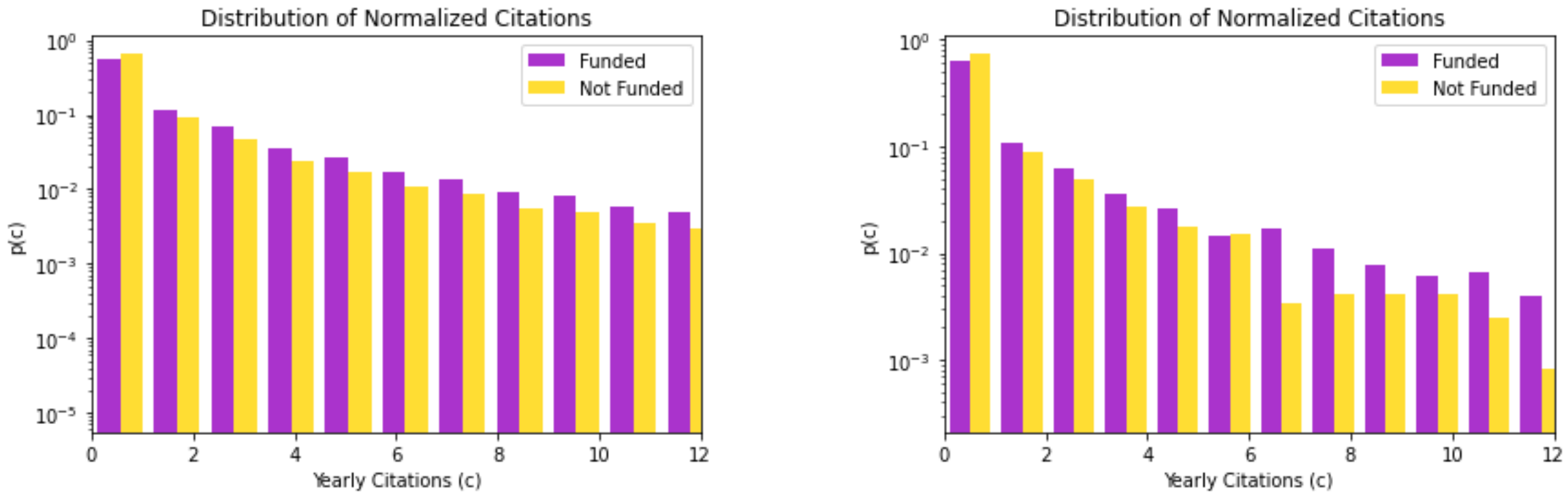

| | |
|---|---|
| Funded : **92905 (63.5%)** Not-Funded : **53423** | Funded : **3166 (73%)** Not-Funded : **1172** |
| Avg. Norm. Citation - Funded : **2.654** | Avg. Norm. Citation - Funded : **2.558** |
| Avg. Norm. Citation - Not Funded : **1.842** | Avg. Norm. Citation - Not Funded : **1.82** |

**Fig. 13. Distribution for normalized citation counts, separately given for funding availability or not. Average values are also provided below the panels.**

Fig.13 yields slightly lower average citation count in the right side, while graph-based studies have noticeably (about 10%) received more funding. In the other hand, for both types of studies, funded studies receive significantly (above 40%) more citation compared to non-funded studies. Moreover, non-funded studies yield greater leftmost bars for the two panels, following smaller bars for the rest of the distribution compared to funded counterparts. This indicates that they have greater probability to receive one citation yearly, while funded studies have slightly greater probabilities for citation counts more than one.

**iv. Publication Type**

Fig. 14 clearly indicates that unrestricted DL studies published in journals receive 18% more citations while graph-based studies published in conference proceedings receive 65% more citations yearly. This outcome also supports our previous inference that graph-based studies distinctly progress through conferences. Also, very close citation performance for journal and conference publications in graph side is not viable for the left side. As in the previous subsection, leftmost conference bars for two panels are greater than journal bars, yielding greater probability for 1 citation for conferences. For both panels, probability of having more than 1 citation is slightly greater for journal publications compared to conference side.

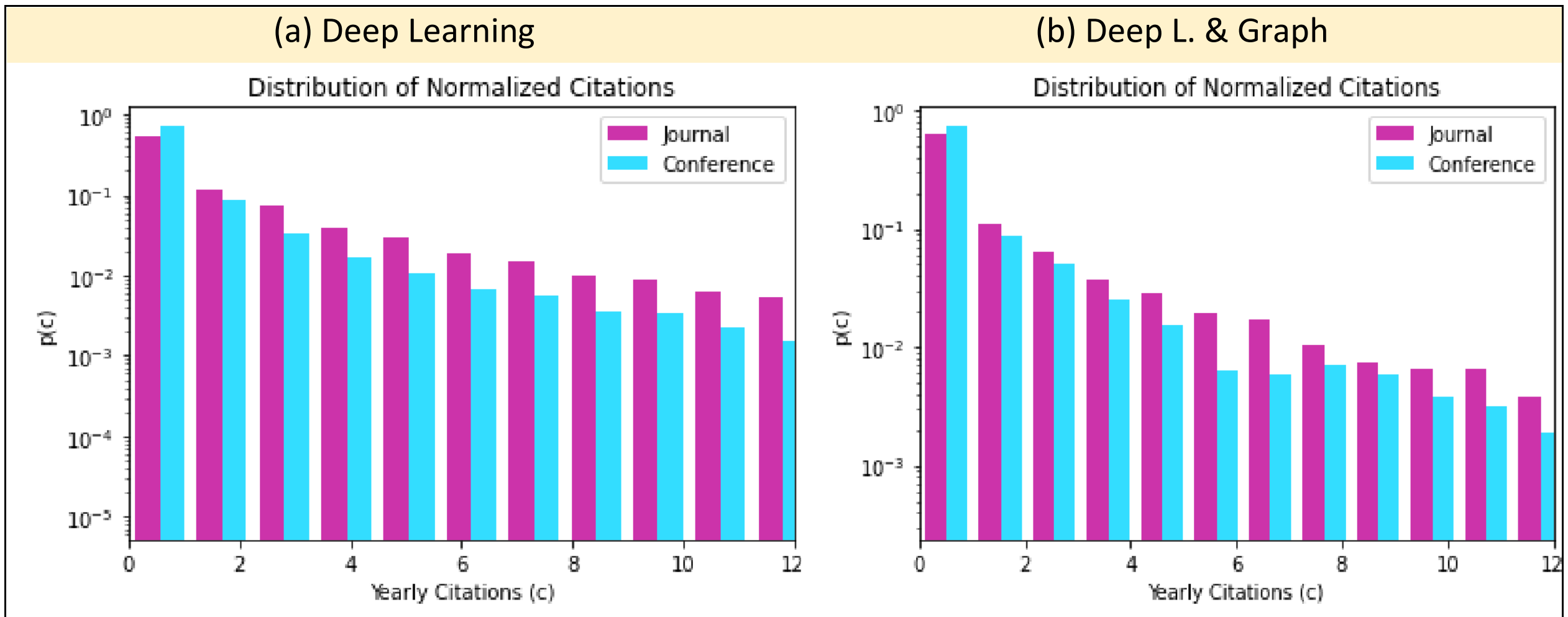

Avg. Norm. Citation - Journal : **2.812** Avg. Norm. Citation - Journal : **2.372**

Avg. Norm. Citation - Conference : **1.416** Avg. Norm. Citation - Conference : **2.341**

**Fig. 14. Distribution for normalized citation counts, separately given for publication type as journal or conference. Average values are also provided below the panels.**

## 4. CONCLUSIONS

Having conducted a bibliometric view on the stream of DL studies and graph-induced DL studies, several inferences emerged as following:

- Graph-based approaches among unrestricted DL stream form approximately 3% of the whole, with a linearly increasing rate which has reached to ~4% nowadays.
- DL itself has an exponentially increasing trend in publication counts.
- Articles (~60%) and proceeding papers (~35%) form the majority of the studies for both streams.
- Some sources tend to publish graph-based studies above average tendency, including IEEE Access, Lecture Notes in Computer Science, Neurocomputing, IEEE Transactions on Geoscience and Remote Sensing, Expert Systems with Applications, IEEE Robotics and Automation Letters. Contrarily, a variety of journals have negative tendency for the same.
- A discriminative view also emerges for research areas, where graph-based approaches are more preferred in biochemistry, computational biology, chemistry, mathematics, computer science and remote sensing fields, generally the most related ones with network science.
- Regional tendencies for graph-based approaches are also observed. Rep. of China, Singapore, Australia, Canada and USA emerge as preferential communities for graph-induced studies, while countries including India, Taiwan, Netherlands, Brazil, Turkey are significantly in negative side.
- Older publications of graph-based studies have significantly greater (~2 times) citation performance compared to unrestricted ones.
- Despite having significantly greater average usage count, graph-based approaches have almost equal average normalized citation to the unrestricted studies. This indicates reading-to-citation convertibility of graph-based studies is visibly lower.

- General DL studies indexed in SCIE or BKCI receive more citation, while CPCI indexed graph-based studies are more cited. Graph-based papers have high impact for conference publications, very close to SCIE-indexed studies. SCIE and CPCI-indexed publications attract 1.5 to 3 times more attention compared to ESCI-indexed publications in DL domain.
- Researchers tend to publish conference papers at a high rate in graph scope, which generally requires less time and effort compared to SCI-E indexed publishing. SSCI, ESCI, AHCI and BKCI indexed studies are strongly in negative side to use graph-based approaches.
- Graph-based studies have 10% more potential to receive funding. Regardless from the graph-enhancement, funded studies receive significantly (above 40%) more citation compared to non-funded studies.
- Unrestricted DL studies published in journals receive 18% more citations while graph-based studies published in conference proceedings receive 65% more citations yearly.

As an overall conclusion, graph-based studies emerge as an increasing sub-trend in DL studies, a stream better progressing through conference events. They need a good understanding to be cited, which is traced in read-to-cite rate and citation boost after some years.